\documentclass[10pt,twocolumn,oneside,letterpaper,conference]{IEEEtran}
\usepackage{cite}
\usepackage{subfigure}
\usepackage{graphicx}
\usepackage{epstopdf}
\usepackage{amsmath}
\usepackage{txfonts}
\usepackage{times}
\usepackage{latexsym}
\usepackage{bm}
\usepackage{amssymb}
\usepackage[center,footnotesize]{caption2}
\usepackage{stfloats}
\usepackage{cases}
\usepackage{array}
\usepackage{setspace}
\usepackage{fancyhdr}
\usepackage{balance} 
\usepackage{flushend} 
\usepackage{multirow}
\usepackage{booktabs}
\usepackage{amsmath}
\usepackage{algorithm}
\usepackage{algpseudocode}
\usepackage{CJK}
\usepackage{xcolor}
\usepackage{amsmath}
\usepackage{booktabs} 
\usepackage{multirow}
\usepackage{array}
\usepackage{bbm}

\allowdisplaybreaks[4]

\newcaptionstyle{one}{%
  \captionfont%
  \onelinecaption%
  {{\centering \captionlabel $.$ \: } 
    \captiontext \par}%
  {{\captionlabel $.$ \: }%
    \captiontext \par}} 
\captionstyle{one}

\begin{document}

\title{\huge Coded Caching via Federated Deep Reinforcement Learning in Fog Radio Access Networks}

\author{
\IEEEauthorblockN{Yingqi Chen$^{1}$,
Yanxiang Jiang$^{1,2,*}$,
Fu-Chun Zheng$^{1,2}$,
Mehdi Bennis$^3$, and Xiaohu You$^1$}
\IEEEauthorblockA{$^1$National Mobile Communications Research Laboratory,
Southeast University, Nanjing 210096, China.\\
$^2$School of Electronic and Information Engineering, Harbin Institute of Technology, Shenzhen 518055, China.\\
$^3$Centre for Wireless Communications, University of Oulu, Oulu 90014, Finland.\\
E-mail: $\{$ chenyingqi@seu.edu.cn, yxjiang@seu.edu.cn, fzheng@ieee.org, mehdi.bennis@oulu.fi, xhyu@seu.edu.cn $\}$
}}

\maketitle

\begin{abstract}
In this paper, the placement strategy design of coded caching in fog-radio access networks (F-RANs) is investigated.
By considering time-variant content popularity, federated deep reinforcement learning is exploited to learn the placement strategy for our coded caching scheme.
Initially, the placement problem is modeled as a Markov decision process (MDP) to capture the popularity variations and minimize the long-term content access delay.
The reformulated sequential decision problem is solved by dueling double deep Q-learning (dueling DDQL).
Then, federated learning is applied to learn the relatively low-dimensional local decision models and aggregate the global decision model,
which alleviates over-consumption of bandwidth resources and avoids direct learning of a complex coded caching decision model with high-dimensional state space. 
Simulation results show that our proposed scheme outperforms the benchmarks in reducing the content access delay, keeping the performance stable, and trading off between the local caching gain and the global multicasting gain.

\end{abstract}

\begin{IEEEkeywords}
Coded caching, federated learning, reinforcement learning, time-variant popularity.
\end{IEEEkeywords}

\section{Introduction}
With the proliferation of smart devices and mobile applications, wireless networks suffer a huge traffic pressure \cite{[1]_review}.
One of the feasible ways to deal with this tremendous data traffic is to transfer the computing tasks and contents to edge devices.
At this point, fog radio access networks (F-RANs) have been proposed \cite{F-RAN} as a promising network architecture.
In F-RANs, fog access points (F-APs) are close to users and endowed with computing capability and storage capacity, which can alleviate network congestion and improve quality of service.

Generally, traditional (uncoded) caching schemes store contents close to users so that the requests can be locally served.
These caching approaches only bring about the \emph{local caching gain}.
In \cite{cen_codedcaching}, a coded caching scheme was proposed to further reduce the fronthaul load by combining caching and multicasting.
In the placement phase, contents are partially stored at cache memories of local users, which provides the \emph{local caching gain}.
In the delivery phase, the server multicasts the coded messages to local users, which provides the \emph{global multicasting gain}.
Based on this work, a large number of modified coded caching schemes have been proposed.

For coded caching with non-uniform content popularity, storing more fractions of a content brings higher local caching gains but loses more global multicasting gain in the meantime. 
In such a case, how to realize the trade-off between the local caching gain and the global multicasting gain is a critical issue.
One of the feasible methods is partitioning the contents into several groups which are handled by different placement and delivery strategies \cite{nonuniform_codedcaching,nonuniform_2,nonuniform_3,multi_request}.
In \cite{nonuniform_codedcaching}, the contents were divided into several groups with similar popularities and each group was cached with different storage space.
In \cite{nonuniform_2} and \cite{nonuniform_3}, the authors demonstrated that dividing contents into two or three groups achieves near-optimal performance.
In \cite{multi_request}, the case of multiple requests was considered and a three-group division strategy was obtained by solving the formulated partition optimization problem.

However, the above coded caching schemes \cite{nonuniform_codedcaching,nonuniform_2,nonuniform_3,multi_request} barely considered time-variant content popularity,
they all designed the placement strategies based on stationary content popularity,
which may degrade the performance in real scenarios.
As for the caching problem with time-variant content popularity, 
some works \cite{RL_cache,RL_cache_1,RL_cache_2} tried to directly made caching decisions by reinforcement learning (RL) without explicit popularity prediction.
In \cite{RL_cache}, the authors modeled maximum-distance separable (MDS) based cooperative caching as a Markov decision process (MDP) to capture the popularity dynamics, and then utilized Q-learning to maximize the long-term expected cumulative traffic load.
In \cite{RL_cache_1}, the authors used direct policy search as the RL algorithm to improve the cache hit rate and demonstrated the importance of feature selection.
In \cite{RL_cache_2}, multi-agent multi-armed bandit based algorithm was proposed to minimize the long-term transmission delay when user preference is unknown.
Nevertheless, these works focused on the non-multicasting caching schemes which omitted the multicasting gain.
An online coded caching scheme without prior knowledge of popularity variations has not been well studied.

Motivated by the aforementioned discussions, we consider the coded caching problem with time-variant content popularity in F-RANs and utilize federated deep reinforcement learning (FDRL) to design the placement strategy.
The placement problem is formulated in the RL framework where dueling double Q-learning (dueling DDQL) is applied as the RL learning algorithm.
The agent is expected to make placement decisions based on the observed dynamic content popularity information to minimize the long-term system transmission delay.
Then, the local placement decision model based on its own local content popularity is independently learned at each F-AP and then aggregated at the cloud server to obtain a suboptimal global model through federated learning.
Through model aggregation, additional costs of the bandwidth resources caused by transmitting training data are reduced
and direct learning of a high-dimensional model is avoided.

The rest of the paper is organized as follows. 
The system model and problem formulation are introduced in Section II.
Section III elaborately presents the FDRL based coded caching scheme.
The simulation results are shown in Section VI.
Finally, Section V concludes the paper.

%
%

%
\section{System Model and Problem Formulation}

\subsection{Network Model}
We consider the F-RANs as illustrated in Fig. \ref{system}. 
The cloud server accesses a library with $N$ contents $ W_1,W_2,\ldots,W_N$, where the size of each content is $F$ bits.
The cloud server is connected with $K$ F-APs through a shared fronthaul link. 
Let $\mathcal{N} = \{1,2,\ldots,N\}$ and $\mathcal{K} = \{1,2,\ldots,K\}$ denote the index sets of contents and F-APs respectively.
Each F-AP with a cache size of $M \times F$ bits serves users via access links.

The considered continuous transmission time is divided into discrete time slots $1,2,\ldots,T$ and the caching updates are made at the end of each time slot.
The local content popularity $p_{k,n}(t)$ represents the popularity of requesting content $n$ under F-AP $k$.
It is noted that the proposed RL based method is designed in a model-free manner so it is valid under any popularity distribution. But for simplicity,
the local content popularity is modeled by the Zipf distribution as follows,
\begin{equation}
  \label{pop}
  p_{k,n}(t) = \frac{1/n^{\alpha_k(t)}}{ {\textstyle \sum_{j=1}^{N}1/j^{\alpha_k(t)}} },
\end{equation} 
where $\alpha_k(t)$ is the distribution parameter which randomly varies with time.
The time-variant local content popularity of F-AP $k$ during time slot $t$ is denoted as $\pmb{p}_k(t)=[p_{k,1}(t),p_{k,2}(t),\ldots,p_{k,N}(t)]^ \mathrm{ T }$.


\subsection{Caching Model}
By considering that each F-AP receives multiple requests from users and individually handles them in the order of arrival,
it is near-optimal to divide the contents into two groups \cite{nonuniform_2}. 
The contents in the first group are cached at F-APs according to the centralized coded caching scheme \cite{cen_codedcaching}.
The contents in the second group are not cached.

\begin{figure}[!t]
  \centering 
  \includegraphics[width=0.42\textwidth]{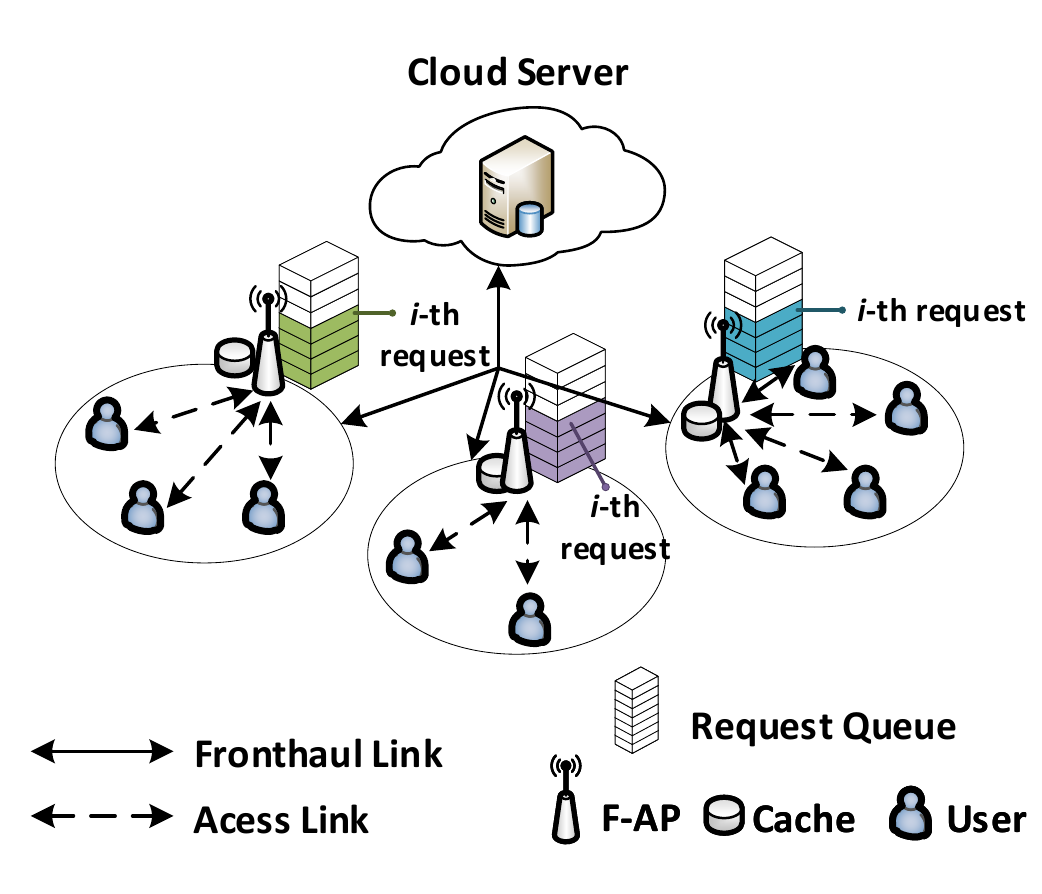}
  \caption{Illustration of the coded caching scenario in F-RANs}
  \label{system}
\end{figure}

In the delivery phase, F-APs queue the received requests as shown in Fig. \ref{system}.
The requests in the same row of all queues are processed together.
For the $i$-th requests of all queues, the number of requests for the contents in the first group is denoted as $u^i$, which are processed by multicasting from the cloud server to the corresponding F-APs,
while the other $K-u^i$ uncached requests are handled by unicasting.
Then, F-APs decode the multicasting messages and send the complete contents to the requesting users.


\subsection{Problem Formulation}
We consider minimizing the total content access delay of all F-APs,
which requires an appropriate placement strategy to specifiy how many and which contents are divided into the first group.

Let $N_c(t)$ denote the number of contents in the first group. 
Before caching, each content in the first group is split into $\binom{K}{L_t}$ fragments \cite{cen_codedcaching}, where $L_t=KM/N_c(t)$.
Let $c_n(t) = 1$ represent that content $n$ is cached, and $c_n(t)=0$ otherwise. The placement strategy during time slot $t$ can be expressed as 
\begin{equation}
  \varPhi(t) =\{c_1(t),c_2(t),\ldots,c_N(t) | {\textstyle \sum_{n=1}^{N}c_n(t)}=N_c(t)\}.
\end{equation}
Let $f_{k,n}^i(t)=1$ represent that the $i$-th request received by F-AP $k$ is content $n$, and $f_{k,n}^i(t)=0$ otherwise.
Then, the number of the cached requests for the $i$-th requests of all queues during time slot $t$ can be expressed as $u^i_t= {\textstyle \sum_{k=1}^{K}}{\textstyle \sum_{n=1}^{N}f_{k,n}^i(t)c_n(t)}$.
In this situation, the cloud server needs to send $\binom{K}{L_t+1}-\binom{K-u^i_t}{L_t+1}$ multicasting messages to F-APs, where the size of each message is $F/\binom{K}{L_t}$ bits.
Specially, when $u^i_t>N_c(t)$, the cloud server can choose to directly multicast the $N_c(t)$ entire contents to the $u^i_t$ F-APs.
Thus, the fronthaul load of the cached requests can be expressed as
\begin{equation}
  \begin{aligned}
    R_c^i(t) &= \min \left( \frac{\binom{K}{L_t+1}-\binom{K-u^i_t}{L_t+1}}{\binom{K}{L_t}}, u^i_t  \left(1-\frac{M}{N_c(t)} \right)  \frac{N_c(t)}{u^i_t} \right) \\
    &=\min \left( \frac{\binom{K}{L_t+1}-\binom{K-u^i_t}{L_t+1}}{\binom{K}{L_t}}, N_c(t) - M \right).
  \end{aligned}
\end{equation}

The remaining $K-u^i_t$ uncached requests require the complete transmission from the cloud server, so the fronthaul load is $R_{uc}^i(t) = K-u^i_t$.
Consequently, the total fronthaul load for the $i$-th row of all queues during time slot $t$ is  
\begin{equation}
  R^i(t) = R_c^i(t) + R_{uc}^i(t) = \min \left( \frac{\binom{K}{L_t+1}-\binom{K-u^i_t}{L_t+1}}{\binom{K}{L_t}}, N_c(t) - M \right) +  (K-u^i_t).
\end{equation}
Finally, F-APs send the requested $K$ contents to the requesting users through access links.

For simplicity, we set the delay of transmitting a complete content through the fronthaul link and the access link as $d_f$ and $d_a$ respectively.
Hence, the total content access delay during time slot $t$ is expressed as 
\begin{equation}
  d(t) = \sum_{i=1}^{V}(d_f \cdot R^i(t) +d_a \cdot K), \label{delay}
\end{equation}
where $V$ denotes the number of the received requests of each F-AP during a time slot. 

Our objective is to find the optimal $N_c^*$ and the corresponding placement strategy $\varPhi^*$ to minimize the total content access delay $d(t)$.
Therefore, the placement optimization problem of the coded caching scheme can be formulated as
\begin{subequations}\label{op}
\begin{align}
  \min_{N_c(t),\varPhi(t)} \quad & d(t) \tag{6}\\
  \mbox{s.t.} \quad
  & M<N_c(t) \leq KM, \quad \label{op_Nc}\\
  & c_n(t) \in \{0,1\}, \quad \forall n \in \cal{N}.
\end{align}
\end{subequations}
For (\ref{op_Nc}), the number of the selected contents $N_c(t)$ should be greater than $M$,
otherwise all contents can be cached.
Besides, for coded caching, keeping $L_t \ge 1$ is essential, so $N_c(t)$ should be no greater than $KM$.

\section{Proposed FDRL based Coded Caching Scheme}
The time-variant content popularity induces a time-varying $d(t)$ in problem (\ref{op}).
Thus, the placement strategy obtained by solving problem (\ref{op}) with traditional schemes will be non-optimal when the content popularity varies.
In order to deal with this issue, an RL framework is proposed to track the popularity variations and model the placement problem as an MDP problem to minimize the long-term content access delay, where dueling DDQL is selected as the RL algorithm.
By considering that the distinct local content popularity among different F-APs makes it difficult to learn the optimal solution directly,
we apply federated learning to learn local coded caching decision model at each F-AP based on its local content popularity and aggregate the local models at the cloud server to obtain a suboptimal global model.

\subsection{Problem Reformulation with RL}
In order to track the time-variant content popularity and intelligently obtain the placement strategy, we resort to RL to model the placement problem (\ref{op}) 
as a MDP \cite{MDP} with unknown transition probability.
The key elements of the RL framework are described as follows.

\subsubsection{State}
At the end of time slot $t$, the state consists of the action taken during time slot $t-1$, i.e., $\pmb{a}\left(t-1\right)$, and the request frequencies for all contents during time slot $t$, i.e., $\bar{\pmb{p}}\left(t\right)$, which is called the statistical content popularity.
The state at time slot $t$ is defined as
\begin{equation}
	\pmb{s}\left(t\right) = \left\{ \pmb{a}\left(t-1\right), \bar{\pmb{p}} \left(t\right) \right\}.
\end{equation}

\subsubsection{Action}

At the end of time slot $t$, the agent determines how many and which contents are cached at F-APs based on the observed state $\pmb{s}(t)$.
When $N$ is relatively small ($N<20$), the action can be described as
\begin{equation}
    \pmb{a} \left(t\right) = \left\{ N_c(t), c_1\left(t\right),c_2\left(t\right),\ldots,c_N\left(t\right) \right\},
\end{equation}
where the dimensionality of the action space increases exponentially with $N$.
When $N$ is relatively large ($N \ge 20$), to reduce the dimensionality, the action can be simplified by choosing the $N_c$ most popular contents according to $\bar{\pmb{p}}_z$, where $\bar{\pmb{p}}_z$ denotes the $z$-th statistical global content popularity that has been observed.


\subsubsection{Transition Probability}
$P\left ( \pmb{s}^{\prime}|\pmb{s},\pmb{a} \right )$ denotes the probability of the state switching from state $\pmb{s}$ to another state $\pmb{s}^{\prime}$ after taking action $\pmb{a}$,
which is unknown to the agent. 
According to \cite{RL_cache}, the time variation of content popularity can be modeled by the transition probability as 
\begin{equation}
  \sum_{j=1}^{Z}P\left([\pmb{a}^{\prime}, \bar{\pmb{p}}_j]\big|[\pmb{a}, \bar{\pmb{p}}_i],\pmb{a}^{\prime}\right)=1, \quad i=1,2,\ldots,Z,
\end{equation}
where $\pmb{a}^{\prime}$ is the action taken under the state $\{\pmb{a},\bar{\pmb{p}}_i\}$.

\subsubsection{Reward}
The environment feeds back the reward $r(t)$ when the state transits from $\pmb{s}(t)$ to $\pmb{s}(t+1)$
and learning from the reward can improve the performance of the agent.
In order to minimize the expected total content access delay, the reward is set according to (\ref{op}) and normalized by the negative exponential function as follows \cite{fd_caching},
\begin{equation} 
    r(t)= \varphi e^{-\sum_{i=1}^{V} (\mu_1 d_f R^i(t)+\mu_2 d_a K) }, \label{reward}
\end{equation}
where $\mu_1$ and $\mu_2$ balance the transmission delays of the fronthaul link and the access link, $\mu_1 + \mu_2 =1$ and $0<\mu_2 < \mu_1 <1$,
and $\varphi$ adjusts the scope of the reward.

\begin{figure}[!t]
  \centering 
  \includegraphics[width=0.35\textwidth]{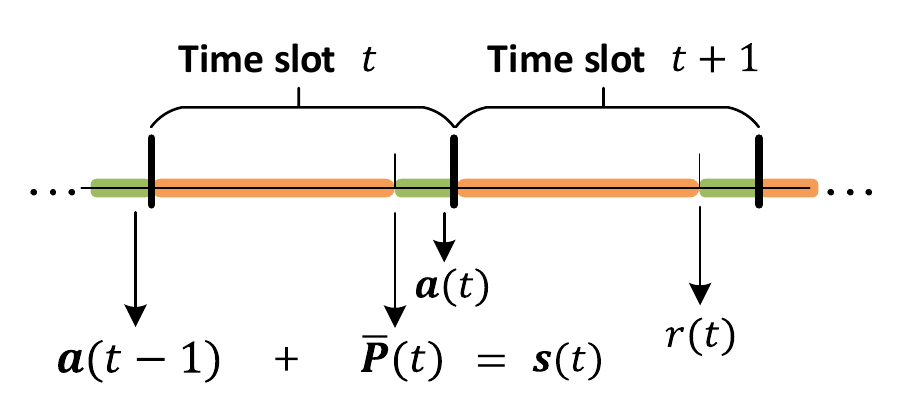}
  \caption{Timeline of the coded caching process} \label{timeslot}
\end{figure}

To provide a clear illustration, the coded caching process is shown in Fig. \ref{timeslot}.
At the end of time slot $t$, the agent observes $\bar{\pmb{p}}\left(t\right)$ which forms $\pmb{s}(t)$ together with $\pmb{a}(t-1)$,
and then the agent makes the caching decision $\pmb{a}(t)$.
At the end of time slot $t+1$, the corresponding reward $r(t)$ is fed back from the environment and the state transits to $\pmb{s}(t+1)$.

Furthermore, the expected discounted cumulative reward can be expressed with the state-action value function \cite{MDP} as
\begin{equation}
    Q_\pi (\pmb{s},\pmb{a}) = \mathbb{E}\left [ \sum_{k=0}^{\infty} \gamma^k r(t+k) | \pmb{s}(t)=\pmb{s},\pmb{a}(t)=\pmb{a}\right ], \label{Q}
\end{equation}
where $\gamma \in [0,1]$ is the discount factor.
Define the policy $\pi$ as a mapping from state $\pmb{s}$ to action $\pmb{a}$.
The goal of RL is to find the optimal policy $\pi^*$ that maximizes $Q_\pi (\pmb{s},\pmb{a})$.

Therefore, in a long term, the problem (\ref{op}) that aims to minimize the delay can be approximately reformulated as the following problem that aims to maximize the long-term expected cumulative reward $Q_\pi (\pmb{s},\pmb{a})$,
\begin{equation}
  \pi^* = \arg \max_{\pmb{a}} Q_\pi (\pmb{s},\pmb{a}). \label{new_target}
\end{equation}
Through continuously interacting with the environment, the agent learns $Q_{\pi}(\pmb{s},\pmb{a})$ which implies the time-varying information about the content popularity profiles,
and then the placement policy can be determined.


\addtolength{\topmargin}{0.01in}
\subsection{Dueling DDQL}

The action space is high-dimensional especially when $N$ is very large.
To tackle this issue, double Q-network \cite{van2016deep} is applied to estimate the Q-value with neural networks, which avoids frequently searching a high-dimensional Q-value table. 
Meanwhile, the dueling architecture \cite{wang2016dueling} is leveraged to accelerate the learning process.

\subsubsection{Double Q-Network}
The actions are chosen according to the online Q-network $Q(\pmb{s},\pmb{a};\theta)$ with parameter $\theta$, while their values are estimated by the target Q-network $\hat{Q}(\pmb{s},\pmb{a};\theta^-)$ with another parameter $\theta^-$. 
Note that only $\theta$ needs to be iteratively updated while $\theta^-$ is duplicated from $\theta$ every $E$ steps.
The periodical updates reduce the correlation between the current Q-value and the target Q-value, therefore mitigating the oscillations and accelerating the convergence.

\subsubsection{Dueling Architecture}
The Q-value function is separately represented by the value function and the advantage function.
The state value function $V(\pmb{s})$ measures the value of a particular state $\pmb{s}$, and the advantage function $A(\pmb{s},\pmb{a})$ indicates the additional value brought by choosing action $\pmb{a}$ in state $\pmb{s}$.
The Q-network with dueling architecture is described as
\begin{equation}
    Q(\pmb{s},\pmb{a};\theta)=V(\pmb{s};\theta^c )+\left(V(\pmb{s},\pmb{a};\theta^v)-\frac{1}{\left | \mathcal{A} \right |  }\sum_{\pmb{a}^{\prime}} A(\pmb{s},\pmb{a}^{\prime};\theta^a )  \right),
\end{equation}
where $\mathcal{A}$ denotes the action space.
The separate representation makes the learning of $Q_{\pi}(\pmb{s},\pmb{a})$ more efficient, thereby accelerating the convergence.


\subsection{FDRL Based Coded Caching}

The coded caching scheme achieves a greater total caching gain than the traditional caching schemes by incorporating local caching and global multicasting.
But the distinct local content popularity among different F-APs complicates the design of the placement strategy
which should consider the local preference and the caching diversity simultaneously to obtain a maximum total caching gain \cite{heter_pop_1,heter_pop_2}.
In such a case, RL agent requires all local content popularity informations to make an optimal placement decision.
As a result, the RL agent should maintain a $((K+1)N+1)$ dimensional state space $\pmb{s}(t)=\left\{ \pmb{a}(t-1), \bar{\pmb{p}}_1(t), \bar{\pmb{p}}_2(t), \ldots, \bar{\pmb{p}}_K(t) \right\}$,
where $\bar{\pmb{p}}_k(t), k \in \mathcal{K}$ is the local statistical content popularity of F-AP $k$.
However, the high-dimensional state space enlarges the number of input features of the Q-network $Q(\pmb{s},\pmb{a};\theta)$,
which increases the complexity of learning and the difficulty of convergence.
Besides, frequently transmitting the training data from F-APs to the cloud server incurs additional costs of bandwidth resources.
To deal with these issues, instead of directly training a model with high dimension, we train the local coded caching decision models that only focuses on the local preference 
and aggregate them to a suboptimal global model at the cloud server in a federated learning way.

In the local model training process, F-AP $k$ designs a selfish local coded caching placement strategy \cite{heter_pop_1} based on its local content popularity and ignores the local preferences of other F-APs.
The local loss function of F-AP $k$ is
\begin{equation}
  L(\theta_k) = \left[ Q_k(\pmb{s}_k,\pmb{a}_k;\theta_k) - Y_k(t) \right]^2 \label{loss},
\end{equation}
where $Y_k(t)$ is the Q-target value and is defined as
\begin{equation}
  Y_k(t) = r_k(t) + \gamma \hat{Q}_k ( \pmb{s}^{\prime}_k, \max_{\pmb{a}^{\prime}_k}Q(\pmb{s}^{\prime}_k,\pmb{a}^{\prime}_k;\theta_k);\theta^-_k).
\end{equation}
After $T_s$ steps of local training, the local model $\theta_k$ is uploaded to the cloud server and aggregated to obtain the improved global model $\theta_G$ which is then distributed to all F-APs \cite{fed}.
Specially, as for the double Q-network, only the network parameter $\theta_k$ of the online Q-network $Q(\pmb{s},\pmb{a};\theta_k)$ is uploaded \cite{federated_DQQN}.
The global loss function is defined as
\begin{equation}
  L_G(\theta_{G}) =  \frac{1}{\textstyle \sum_{k \in \cal{K}}D_k} \sum_{k=1}^{K}D_kL(\theta_k) ,
\end{equation}
where $D_k$ is the local training data size of F-AP $k$. The aggregation is expressed as
\begin{equation} \label{aggregation}
  \theta_{G} = \frac{1}{{\textstyle \sum_{k \in \cal{K}}}D_k} \sum_{k=1}^{K} D_k \theta_k.
\end{equation}
In such a model integration manner, the local state of F-AP $k$ is defined as $\pmb{s}_k(t)=\left\{ \pmb{a}_k(t-1), \bar{\pmb{p}}_k(t) \right\}$, where $\pmb{a}_k(t)$ is the local action of F-AP $k$.
As a consequence, the dimension of the state is decreased to $2N+1$, which reduces the learning complexity.
\begin{algorithm}[!t]
  \caption{Training Process at Local F-AP}
  \label{local training}
  \begin{algorithmic}[1]
  \Require
  \Statex The experience memory $\mathcal{D}_k$ with capacity $B$;
  \Statex Batch size $b$ and learning rate $\varepsilon$;
  \Statex The $Q$ network with parameter $\theta_k$;
  \Statex The $\hat{Q}$ network with parameter $\theta_k^- = \theta_k$;
  \Ensure
  \For {time slot $t=1,2,\ldots,T_s$}
    \State Observe state $\pmb{s}_k(t)$;
    \State Randomly choose action with probability $\epsilon$, otherwise select $\pmb{a}_k(t)=\arg \max_{\pmb{a}_k(t)}=Q(\pmb{s}_k(t),\pmb{a}_k(t);\theta_k)$;
    \State F-AP $k$ receives $V$ requests from users and evenly divides the requests into $K$ parts;
    \State Calculate the reward $r_k(t)$ according to (\ref{reward});
    \State State transits to $\pmb{s}_k(t+1)$;
    \State Save $[\pmb{s}_k(t),\pmb{a}_k(t),r_k(t),\pmb{s}_k(t+1)]^\mathrm{T}$ in $\mathcal{D}_k$;
    \State Randomly sample $b$ experiences from $\mathcal{D}_k$;
    \State Update $\theta_k$ by gradient descent on (\ref{loss}) with $\varepsilon$;
    \State Replace $\theta_k^-$ with $\theta_k$ every $E$ steps;
  \EndFor
  \end{algorithmic}
  \end{algorithm}

Nevertheless, single each F-AP is unable to implement its coded caching policy. 
To solve this issue, virtual coded caching is implemented at each F-AP.
In the local training process, F-AP $k$ selects an action $\pmb{a}_k(t)$ according to the local model $\theta_k$ based on the observed state $\pmb{s}_k(t)$ but does not actually perform it.
At the end of time slot $t$, the received requests are evenly divided into $K$ parts, which can be regarded as the requests received by the $K$ F-APs with the same local content popularity $\pmb{p}_k(t)$.
Then, the reward $r_k(t)$ can be calculated according to (\ref{reward}) with the known delays $d_a$ and $d_f$.
Thus the experience $[\pmb{s}_k(t),\pmb{a}_k(t),r_k(t),\pmb{s}_k(t+1)]^\mathrm{T}$ can be collected for training. 
In practice, F-APs perform the coded caching decisions $\pmb{a}_G(t)$ according to the global model $\theta_G$ 
based on the observed global state $\pmb{s}_G(t)=\left\{ \pmb{a}_G(t), \bar{\pmb{p}}_G(t) \right\}$, where $\bar{\pmb{p}}_G(t)$ is the statistical global content popularity.
The local training process is shown in Algorithm \ref{local training},
and the complete FDRL based coded caching scheme is shown in Algorithm \ref{global training}.

\addtolength{\topmargin}{0.01in}
\section{Simulation Results}

\begin{algorithm}[!t]
  \normalsize
  \caption{FDRL Based Coded Caching Scheme}
  \label{global training}
  \begin{algorithmic}[1]
  \Require
  \Statex The global model with random $\theta_G$;
  \Statex The local model of each F-AP with $\theta_k=\theta_k^{-}=\theta_G$;
  \Statex The aggregation step $T_s$;
  \Ensure
  \For{time slot $t=1,2,\ldots,T$}
    \State Select $\pmb{a}(t)=\arg \max_{\pmb{a}(t)}=Q(\pmb{s}(t),\pmb{a}(t);\theta_G)$ and perform the corresponding caching;
    \State Each F-AP executes Algorithm \ref{local training} simultaneously;
    \If{$ t \mod T_s=0 $}
        \State Each F-AP sends $\theta_k$ to the cloud server;
        \State Update $\theta_G$ according to (\ref{aggregation});
        \State Distribute $\theta_G$ to all F-APs;
        \State Each F-AP sets $\theta_k=\theta_k^{-}=\theta_G$;
    \EndIf
  \EndFor
  \end{algorithmic}
  \end{algorithm}


The performance of the proposed FDRL based coded caching scheme is evaluated via simulations.
The parameters are set as: $\alpha_k(t)\in [0.5,1.5]$, $N=200$, $Z=10$, $V=50$, $d_f=5$ms, $d_a=1$ms, $\mu_1=0.95$, $\mu_2=0.05$, $\varphi =3$, $B=5000$, $b=32$ and $\varepsilon=0.001$.
Besides, the step interval of updating $\theta_k^-$ to $\theta_k$ is $T_{\text{update}} = 200$, which is bigger than the aggreation step $T_s=20$ \cite{federated_DQQN}.

Four benchmarks are selected for comparison: Arbitrary Probability Coded Caching (APCC) \cite{nonuniform_2}, Non-Uniform Coded Caching (NUCC) \cite{multi_request} and Least Frequently Used (LFU).
The centralized scheme deploys only one agent at the cloud server and make placement decisions according to the observed global statistical content popularity. 
APCC partitions contents whose popularity is greater than a fixed value into the first group. 
NUCC obtains the partition strategy by solving an optimization problem. 
LFU completely stores the $M$ most popular contents.

\begin{figure}[!t]
  \centering 
  \includegraphics[width=0.40\textwidth]{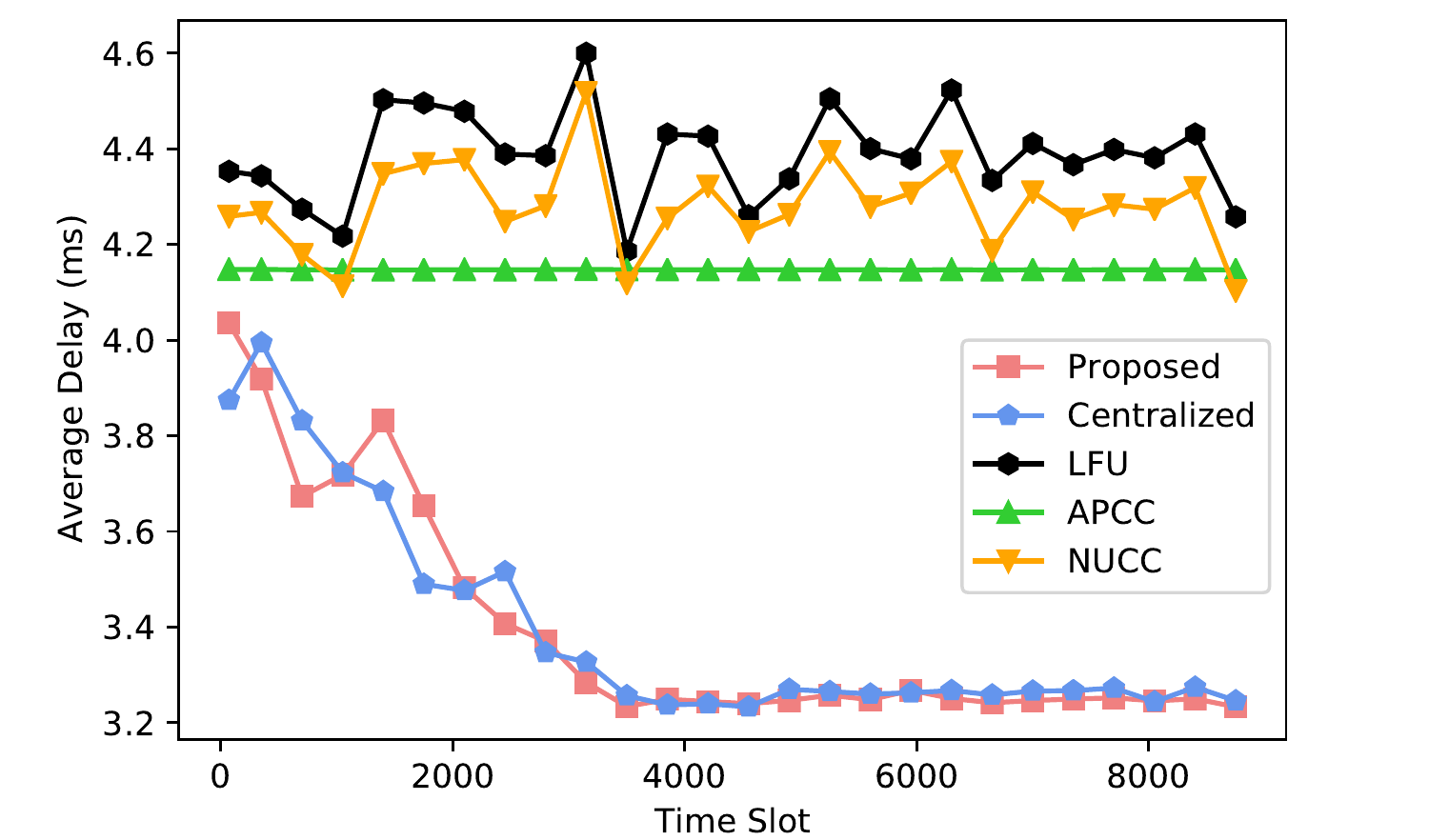} 
  \caption{Performance evaluation in terms of the average delay with respect to time.}  \label{result1_1}
\end{figure}

In Fig. \ref{result1_1}, the performance of the average delay with $K=5$ and $M=30$ is shown.
It can be observed that the proposed scheme achieves the lowest delay and remains relatively stable after convergence.
which is similar to the centralized scheme.
The reason is that the proposed scheme captures the popularity variation and makes the caching decisions adapting to the dynamic environment.
The average delay of NUCC and LFU fluctuates due to neglect of the time varying information.
The average delay of APCC keeps steady because it partitions most of the contents into the first group.
The observations verify that the proposed scheme is able to achieve the stable low-delay transmissions in a federated learning manner.

\begin{figure}[!t]
  \centering 
  \includegraphics[width=0.40\textwidth]{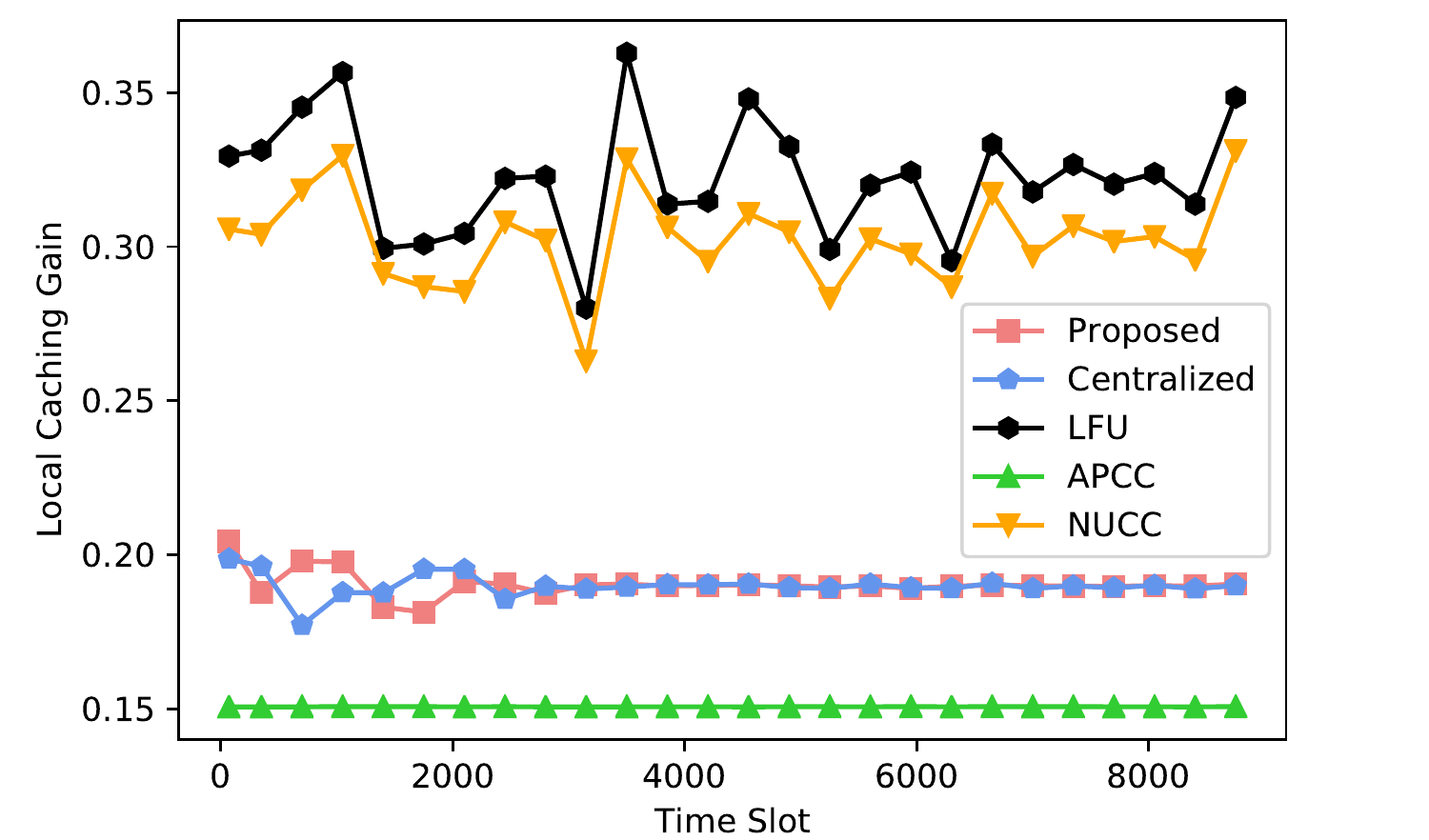} 
  \caption{Performance evaluation in terms of the local caching gain with respect to time.} \label{result1_2}
\end{figure}

In Fig. \ref{result1_2},  the performance of the local caching gain with $K=5$ and $M=30$ is shown. 
The metric \emph{local caching gain} is defined as the hit rate multiplied by the caching fraction of content.
Larger local caching gain means more fractions of content are stored, which also means more multicasting opportunities are vanished. 
It can be observed that the local caching gain of LFU and NUCC is greater than that of the proposed scheme, and APCC is the minimum. 
The reasons are that LFU does not multicast and it only has the local caching gain, while the other schemes trade the global multicasting gain with the local caching gain in various degrees.
Although NUCC has larger local caching gain, it achieves the higher delay due to the loss of multicasting opportunities.
On the contrary, APCC caches most of the contents to catch more multicasting opportunities while losing the local caching gain.
The proposed scheme has a moderate local caching gain and achieves the lowest delay, which indicates that the proposed scheme surpasses the other schemes in trading off between the local caching gain and the global multicasting gain.

\begin{figure}[!t]
  \centering 
  \includegraphics[width=0.40\textwidth]{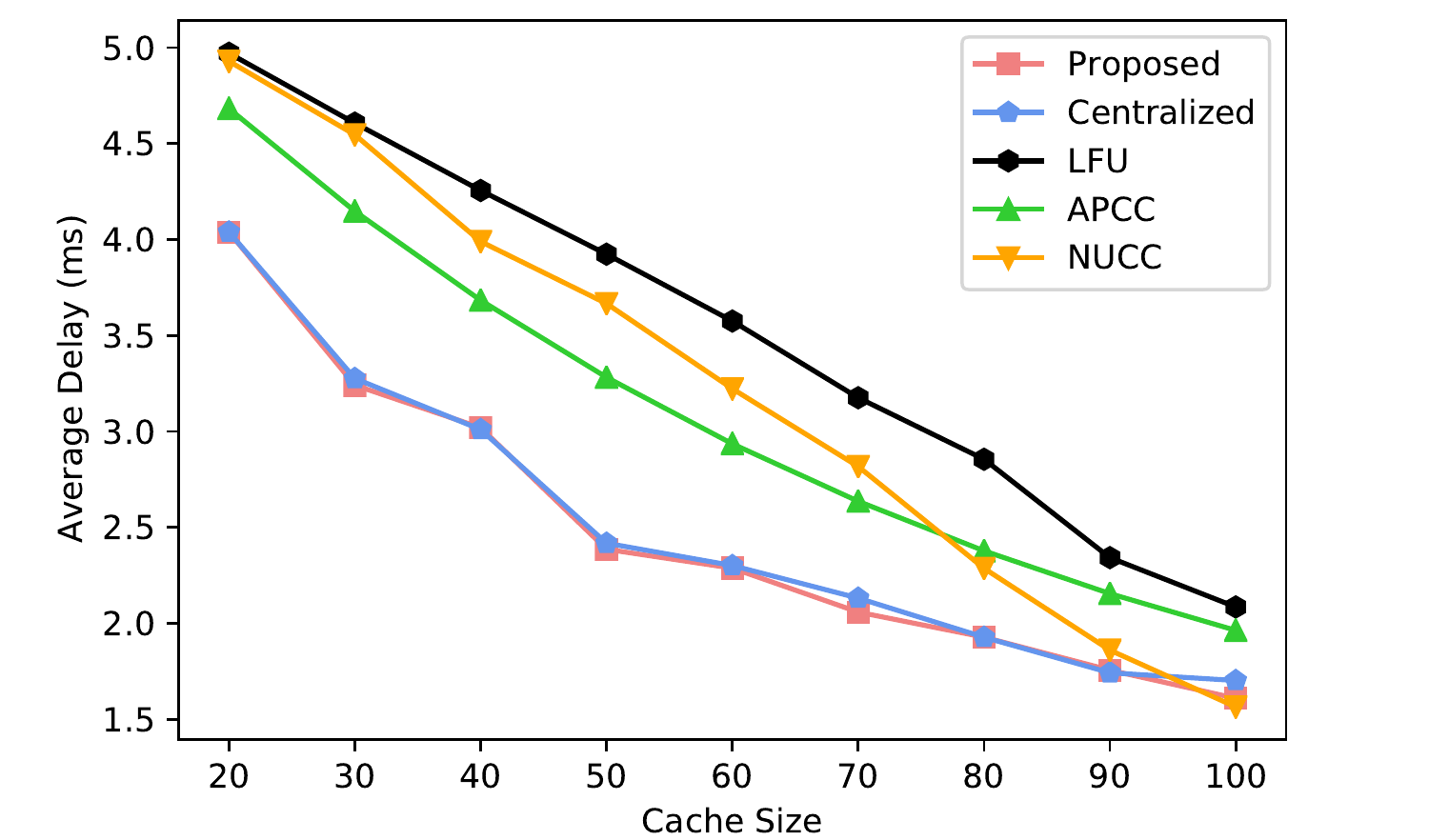}
  \caption{Performance evaluation in terms of the average delay with respect to cache size} \label{result2}
\end{figure}

In Fig. \ref{result2}, we show the performance under different cache size $M$ with $K=5$.
The proposed scheme has the lowest delay and the performance gap becomes smaller with the increase of $M$.
It is also observed that the average delay of each scheme is reduced with the increase of $M$, which is because the bigger caching capacity enables F-APs to store more contents.

\begin{figure}[!t]
  \centering 
  \includegraphics[width=0.40\textwidth]{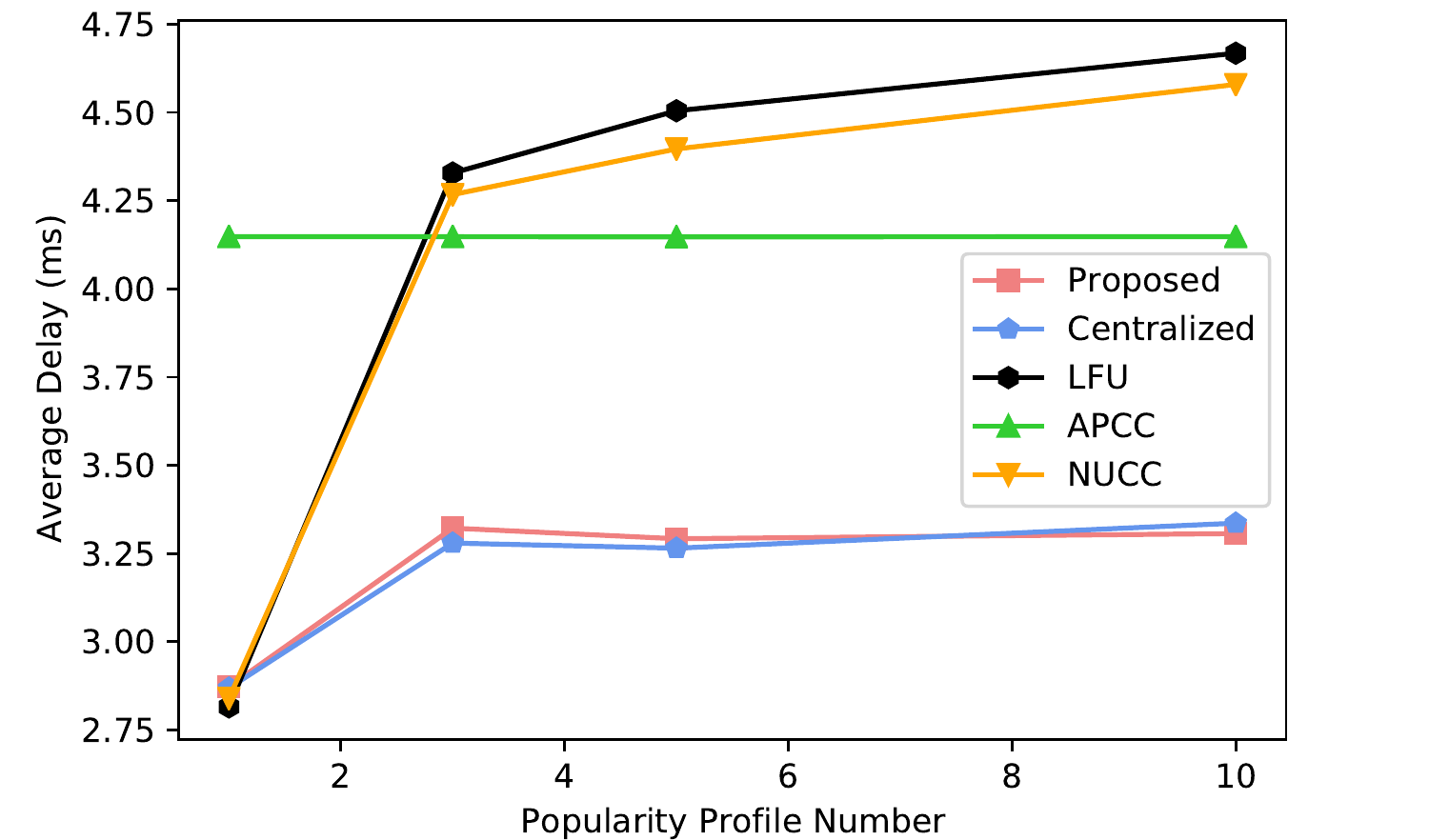}
  \caption{Performance evaluation in terms of the average delay with respect to popularity profile number} \label{result3}
\end{figure}


In Fig. \ref{result3}, the performance under different numbers of popularity profiles $Z$ with $K=5$ and $M=30$ is shown.
The popularity variation becomes more complex with increasing $Z$.
It can be observed that the delay of the proposed scheme goes up when $Z$ increases from 1 to 3 and then remains stable,
while the delays of LFU and NUCC keep rising with the increase of $Z$.
The reason is that the proposed scheme utilizes FDRL to track the popularity variation,
hence the performance is slightly affected.
Instead, LFU and NUCC ignore the variation.
The delay of APCC stays constant because it always stores most of the contents.

\section{Conclusions}
In this paper, we have proposed the FDRL approach to design the coded caching scheme in the F-RANs with time-variant content popularity.
By formulating the placement problem as an MDP and exploiting the advantages of the FDRL,
the placement strategy is obtained by cooperatively learning a shared predictive model at F-APs.
Simulation results have shown that the RL framework is able to tackle the time-variant popularity and the proposed scheme achieves the stable low-delay transmission.
Furthermore, the simulation results have also revealed that for the coded caching scheme, focusing only on the local caching gain or the global multicasting gain is non-optimal,
and ignoring the content popularity variation is detrimental to the actual performance.

\section*{Acknowledgments}
This work was supported in part by the National Key Research and Development Program under Grant 2021YFB2900300,
the National Natural Science Foundation of China under grant 61971129,
and the Shenzhen Science and Technology Program under Grant KQTD20190929172545139.

\bibliographystyle{IEEEtran}
\bibliography{reference}
\end{document}